\documentclass[ 10pt, prl, reprint, superscriptaddress, aps ]{revtex4-2}
\usepackage{textcomp} 
\newcommand{\qdist}[1]{\ifmmode\langle#1\rangle\else\textlangle#1\textrangle\fi}
\usepackage{svg}
\usepackage{lipsum}
\usepackage{amsmath}
\usepackage{empheq}
\usepackage{tikz}
\usepackage{amsthm}
\usepackage{float}
\usepackage{amssymb}
\usepackage{comment}
\usepackage{graphicx}
\usepackage{dcolumn}
\usepackage{bm}
\usepackage{braket}
\usepackage{physics}
\usepackage[hidelinks]{hyperref}
\usepackage{cleveref}
\usepackage{silence}
\begin{document}

\title{Number-correlated photon entanglement with a quantum memory interferometer}

\author{Imogen Forbes}
\affiliation{Quantum Engineering Technology Laboratories, H. H. Wills Physics Laboratory and School of Electrical,
Electronic, \& Mechanical Engineering, University of Bristol, BS8 1FD, UK}

\author{Vincent Boyer}
\affiliation{Department of Physics \& Astronomy, University of Birmingham,  Edgbaston, Birmingham, B15 2TT, UK.}

\author{Louis Edwards-Pratt} 
\affiliation{Department of Physics \& Astronomy, University of Birmingham,  Edgbaston, Birmingham, B15 2TT, UK.}

\author{Mahirul Khan} 
\affiliation{Department of Physics \& Astronomy, University of Birmingham,  Edgbaston, Birmingham, B15 2TT, UK.}

\author{Sougato Bose}
\affiliation{Department of Physics \& Astronomy, University College London, Gower Street, WC1E 6BT London, UK}

\author{Thomas M Hird}
\email{t.m.hird@bham.ac.uk}
\affiliation{Department of Physics \& Astronomy, University of Birmingham,  Edgbaston, Birmingham, B15 2TT, UK.}

\date{\today}

\begin{abstract}
Strong photon–photon interactions are central to quantum optical technologies yet remain challenging to realise at the single-photon level. Here, we show how embedding a quantum memory within a Mach-Zehnder interferometer can offer an alternative to strong nonlinerities, and enable a coherent transformation that converts single-mode photon states into number-correlated entangled states. Using an off-resonant ladder-type atomic interaction, storage and retrieval within a quantum memory maps a single-photon excitation onto a pair of photons across distinct optical modes via a driving control field. Embedding this within an interferometer deterministically produces an effective transformation in the photon number basis: $\sum_n a_n\ket{n} \rightarrow\sum_n a_n\ket{n}\ket{n}$ without relying on conventional optical nonlinearities. Measurement of one mode enables photonic non-demolition detection, state preparation, and frequency conversion as applications of the same underlying transformation. These results establish quantum memories embedded within an interferometric architecture as a platform for photon-photon interactions and the generation of Gaussian and non-Gaussian entangled states.
\end{abstract}

\maketitle

\section{Introduction}

Single photons are central to many quantum technologies, but their weak interactions limit the implementation of key processes such as non-demolition detection and photon-number dependent operations. Overcoming this limitation is a central challenge in quantum optics.

A particularly important capability is quantum non-demolition (QND) measurement of photon number. Existing approaches typically rely on strong optical nonlinearities \cite{balybin2022quantum} or cavity QED \cite{nogues1999seeing,guerlin2007progressive}, which impose stringent requirements on coherence, light-matter coupling strength, mode volume tightness and mode matching, and remain challenging to scale \cite{yanagimoto2023quantum}.

Quantum memories provide a natural light–matter interface, enabling reversible mapping of photonic states onto collective excitations. Their demonstrated ability to store and retrieve weak optical fields/photonic quantum states while preserving photon-number statistics suggests that they can mediate interactions without degrading bosonic excitation structure \cite{heshami2016quantum}.

Here we show that embedding a quantum memory within an optical interferometer enables a conditional photon-addition process that generates number correlations between distinct optical modes. This arises from phase-matched retrieval of stored excitations into a well-defined control mode, combined with interferometric mode selection.

This mechanism produces an effective photon-number–correlating transformation and enables applications including non-destructive photon detection, state preparation, and frequency conversion.

\section{Anti-symmetric photon addition to a balanced Mach-Zehnder Interferometer} 

In this section we consider a balanced Mach–Zehnder interferometer (MZI) with a coherent-state input, \(\ket{\alpha}\), entering port one and vacuum, \(\ket{0}\), entering port two. Here, we label the spatial modes between beamsplitters independently: input modes $A$ and $B$, MZI arm modes $u$ and $l$, and output modes $C$ and $D$. The first 50:50 beamsplitter transforms the input modes to the modes of the MZI arms such that
\begin{equation}
\hat{A}^{\dagger} \rightarrow \frac{\hat{u}^{\dagger} + \hat{l}^{\dagger}}{\sqrt{2}}, \qquad
\hat{B}^{\dagger} \rightarrow \frac{\hat{u}^{\dagger} - \hat{l}^{\dagger}}{\sqrt{2}}.
\end{equation}
For the input state $\ket{\alpha}_A\ket{0}_B$, the state within the arms of the MZI becomes
\begin{equation}
\ket{\psi_{\text{MZI}}}
= \ket{\tfrac{\alpha}{\sqrt{2}}}_u
  \ket{\tfrac{\alpha}{\sqrt{2}}}_l.
\label{eqn:mzi_state}
\end{equation}
The operator transformations of arm modes to output modes peformed by the second beam splitter are given by
\begin{equation}
\frac{\hat{u}^{\dagger} + \hat{l}^{\dagger}}{\sqrt{2}} \rightarrow \hat{C}^{\dagger},\qquad
\frac{\hat{u}^{\dagger} - \hat{l}^{\dagger}}{\sqrt{2}} \rightarrow \hat{D}^{\dagger},
\label{eqn:dark_port}
\end{equation}
such that output state of the MZI is 
\begin{equation}
\ket{\psi_{\text{out}}} = \ket{\alpha}_C\ket{0}_D.
\label{eqn:out_state}
\end{equation}
Hence the probability $P_\mathrm{Dark}$ (excluding imperfections and noise) of a detection event at the dark port $D$ is $P_\mathrm{Dark}=0$, as expected.

We now consider the scenario of adding $n$ photons to the state within the arms of the MZI defined in Eq.~(\ref{eqn:mzi_state}). In particular, we consider the case where these photons are in the specific anti-symmetric superposition across the arms $\frac{(u^{\dagger} - l^{\dagger})}{\sqrt{2}}$. Operationally, for $n$-photons this corresponds to applying {$\left[\frac{\hat{u}^\dagger-\hat{l}^\dagger}{\sqrt{2}}\right]^n$} to the state \(\ket{\psi_\text{MZI}}\). The state within the arms of the MZI therefore becomes
\begin{equation}
\ket{\psi_{\rm  MZI}^{(n)}}
\propto
\left[\frac{\hat{u}^\dagger-\hat{l}^\dagger}{\sqrt{2}}\right]^n \ket{\frac{\alpha}{\sqrt{2}}}_u
\ket{\frac{\alpha}{\sqrt{2}}}_l.
\label{eqn:mzi_state_n}
\end{equation}
This form is advantageous as, in reference to Eq.~(\ref{eqn:dark_port}), the anti-symmetric superposition will produce an $n$-photon output at the dark port, thus producing a detectable signature of photon addition. Utilising the previous result of Eq.~(\ref{eqn:out_state}), the new photon-added output state following the second beamsplitter will become 
\begin{align}
\ket{\psi^{(n)}_{\text{out}}} &= (\hat{D}^{\dagger})^n\ket{\alpha}_C\ket{0}_D \\
&= \ket{\alpha}_C\ket{n}_D,
\end{align}
producing an $n$-photon signal at the dark port. The photon number measured at the dark port therefore provides a means to identify both the presence and extent of photon addition in the arms of the MZI.

\section{Quantum memory generation of photonic excitations}

\begin{figure}[b]
    \centering
    \includegraphics[width=1\linewidth]{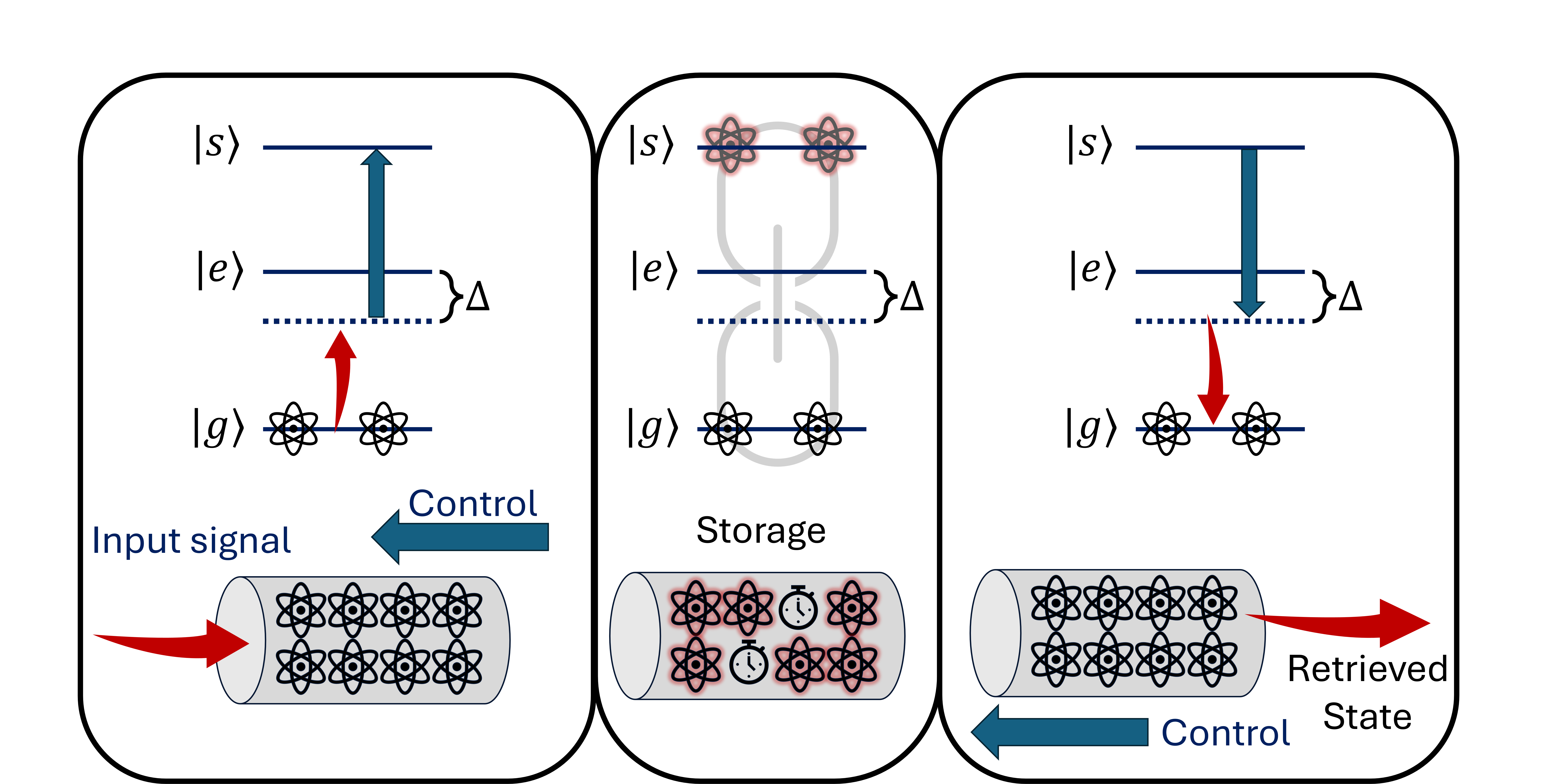}
    \caption{Energy-level diagram for a ladder quantum memory, where an incoming signal is stored as a delocalised excitation across an atom ensemble before being re-emitted following a second control pulse being applied to the ensemble.}
    \label{fig:levels}
\end{figure}

In this section, we show that a quantum memory interaction can facilitate the photon addition operation described above. We begin by considering the interaction between an optical signal mode, a collective spin-wave excitation in an atomic quantum memory, and a driving control field. The interaction can be modelled by the standard trilinear three-wave-mixing Hamiltonian \cite{walls1995quantum}, adapted here to the signal, spin-wave and control modes of a ladder quantum memory 
\begin{equation}
\hat{H} = \hbar g \left( \hat{s}\, \hat{b}^\dagger \hat{c} + \hat{s}^\dagger \hat{b}\, \hat{c}^\dagger \right),
\label{eq:Hamiltonian}
\end{equation}
where $\hat{s}$ denotes the annihilation operator for the signal mode, $\hat{b}$ the annihilation operator for the collective spin-wave excitation, $\hat{c}$ the annihilation operator for the control field, and $g$ a coupling constant. This interaction describes ladder-type quantum memory protocols \cite{kaczmarek2018high}, where optical excitations are coherently mapped onto and retrieved from a collective, doubly excited, atomic degree of freedom. This is shown in Fig~\ref{fig:levels}.

The first term of Eq.~(\ref{eq:Hamiltonian}) describes the storage process, in which a signal photon and a control photon are annihilated to create a collective excitation. The second term describes the retrieval process, in which a stored excitation is converted back into an optical signal mode while simultaneously generating an excitation in a second control field.

These ladder memories have been shown to operate without saturation and preserve photon-number statistics, as demonstrated experimentally for both single photons and weak coherent states \cite{kaczmarek2018high,kaczmarek2017orca}. This confirms that the storage and retrieval processes act linearly on bosonic excitations. These excitations undergo phase-matched retrieval into a well-defined control mode, ensuring the retrieved control excitation is emitted into the same spatio-temporal mode occupied by the retrieval control pulse. As such the interaction produces a photon-added state, conditional on successful retrieval.

Importantly, the ladder-type memory interaction is shown to be functionally noise-free (including being absent from four-wave mixing noise that can occur in $\Lambda$-type Raman memories \cite{kimiaee2025unwanted,thomas2019raman}), whilst the phase-matching conditions prevent spontaneous noise sources. Furthermore, when the control beam is applied to the upper rung of the ladder transition, scattering processes for that beam can be eliminated. Spin-wave decoherence, governed by Doppler dephasing, manifests as a reduction in retrieval efficiency rather than as a source of added noise photons \cite{kaczmarek2017orca}. Consequently, imperfections in the memory affect only the success probability of the protocol, not the fidelity of the conditional output state.

\subsection{Memory-induced number shift} 

The retrieval process can be understood directly from the conservation law associated with the interaction Hamiltonian. Before retrieval, the signal mode is empty and the stored excitation resides in the memory. The retrieval interaction conserves the quantity 
\begin{equation}
\hat Q = \hat N_s - \hat N_c , 
\end{equation} 
where $\hat N_s=\hat s^\dagger \hat s$ and $\hat N_c=\hat c^\dagger \hat c$ are the signal and control photon-number operators respectively. Since 
\begin{equation} 
[\hat H,\hat Q]=0, 
\end{equation} 
the difference in photon number between the signal and control modes remains constant throughout retrieval. Consider a stored $n$-photon excitation and a control pulse initially containing $N$ photons, 
\begin{equation} 
|0\rangle_s |n\rangle_b |N\rangle_c . 
\end{equation} 
Complete retrieval converts the spin-wave excitation into $n$ signal photons, 
\begin{equation} 
|0\rangle_s |n\rangle_b |N\rangle_c \rightarrow |n\rangle_s |0\rangle_b |N+n\rangle_c . 
\end{equation} 
Each retrieved signal photon is therefore accompanied by the addition of exactly one photon to the control mode. It is convenient to define the number-shift operator 
\begin{equation}
\hat S_n = \sum_{N=0}^{\infty} |N+n\rangle\langle N|, \label{eq:shiftoperator} \end{equation} 
whose action is 
\begin{equation} \hat S_n |N\rangle = |N+n\rangle . \end{equation} 

For an arbitrary control state $|\phi\rangle_c$, retrieval implements the transformation 
\begin{equation} 
|n\rangle_b |\phi\rangle_c \rightarrow |n\rangle_s \hat S_n |\phi\rangle_c . \label{eq:retrieval_shift} 
\end{equation} 
Consequently, a stored state 
\begin{equation}
|\psi\rangle_b = \sum_{n=0}^{\infty} a_n |n\rangle_b 
\end{equation} 
evolves according to 
\begin{equation} |\psi\rangle_b |\phi\rangle_c \rightarrow \sum_{n=0}^{\infty} a_n |n\rangle_s \hat S_n |\phi\rangle_c . \label{eq:full_mapping} 
\end{equation} 

Equation~(\ref{eq:full_mapping}) is the fundamental memory-induced transformation and establishes entanglement between the retrieved signal and control fields for arbitrary input states.

For the remainder of this work, we consider a strong coherent retrieval pulse, $|\phi\rangle_c = |\alpha\rangle$, with $|\alpha|^2 \gg n$. In this regime, the addition of $n$ photons produces a small excitation on top of a large classical background, and the shifted coherent state 
\begin{equation} \hat S_n |\alpha\rangle = e^{-|\alpha|^2/2} \sum_{N=0}^{\infty} \frac{\alpha^N}{\sqrt{N!}} |N+n\rangle 
\label{eq:shifted_coherent_state}
\end{equation} approaches the state of a creation operator acting on a coherent state. We therefore approximate the action of $\hat S_n$ to the addition of $n$ photons to the coherent mode as, 
\begin{equation} \hat S_n |\alpha\rangle \approx (\hat c^\dagger)^n |\alpha\rangle . \label{eq:large_alpha_approx} 
\end{equation} 
We validate this approximation in the appendix, taking the fidelity between the two states and showing that this scales as $\mathcal{F}=1-\frac{n^2}{4|\alpha|^2}$ to first order in $1/\alpha^2$.

This approximation allows the interferometric analysis below to be formulated in terms of photon-added coherent states while retaining the exact number-conserving description of Eq.~(\ref{eq:full_mapping}).

\section{Quantum Memory Interferometer}

\begin{figure*}[ht]
    \centering
    \includegraphics[width=1\linewidth]{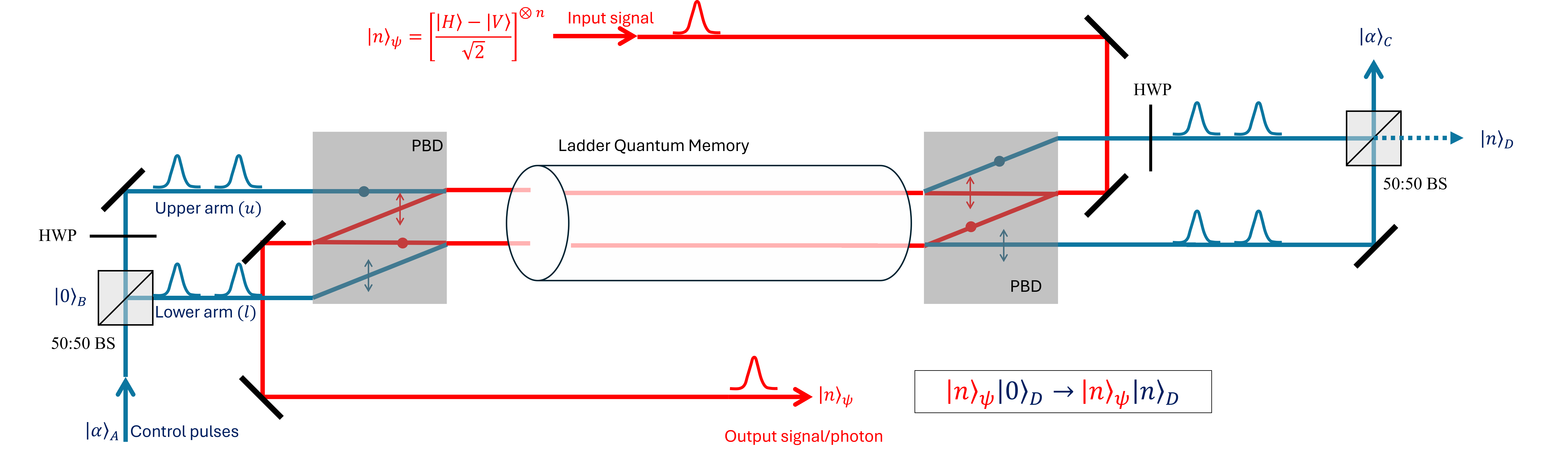}
    \caption{The Quantum Memory Interferometer (QMI). An input signal photon prepared in the $|{-}\rangle = (|H\rangle - |V\rangle)/\sqrt{2}$ polarisation state is split into two spatial paths by a Polarisation Beam Displacer (PBD), travels through the atoms, and is stored and retrieved by the quantum memory process. Counterpropagating to the signal is a control beam comprising of two pulses separated in time. The control pulses are split into two orthogonal polarisations by a 50:50 beamsplitter and a HWP. In the case when storage and retrieval has occurred in both arms, each of the control pulses will have an extra $n$ photons added. Due to the anti-symmetric nature of the signal and this addition, when these control beams interfere at the second 50:50 beamsplitter, they produce an $ n$-photon Fock state at the dark port of the MZI.}
    \label{fig:ORCA1}
\end{figure*}

In the previous two sections, we have shown that:
\begin{enumerate}
    \item If we can add (an anti-symmetric superposition of) $n$ photonic excitations within a balanced MZI of coherent states, then we can interferometrically isolate those added excitations and retrieve the well-defined Fock state $\ket{n}$ from the MZI's dark port.
    \item A ladder quantum memory can (conditional on successful storage and retrieval) add $n$ photonic excitations to the control pulse used to retrieve a $\ket{n}$ signal photon state.
\end{enumerate}

We now combine these results, as shown in Fig. \ref{fig:ORCA1}, into a device we term a \textit{Quantum Memory Interferometer} (QMI). For the purposes of this derivation, we shall assume unit read-in and read-out efficiency, before considering imperfect efficiency in the following section. This configuration utilises a dual-rail quantum memory configuration that can be used to store polarisation qubits in single-mode quantum memories \cite{england2012high,gera2024hong}. Between the two 50:50 beamsplitters, the control beam is in a balanced MZI configuration, with both paths passing through the atoms. 

Let us consider an $n$-photon $\ket{n}$ signal prepared in the $|{-}\rangle = (|H\rangle - |V\rangle)/\sqrt{2}$ polarisation state. The polarisation beam displacer (PBD) maps orthogonal polarisations to distinct spatial paths, so that $|{-}\rangle$ maps onto the antisymmetric spatial superposition $(\ket{u} - \ket{l})/\sqrt{2}$ required by the interferometric scheme above. If the first pulse of the control is overlapped with the signal inside the atoms, then the memory interaction occurs, and the signal is stored in a spinwave across the atomic ensemble. 

When the second control pulse is incident on the atoms, the signal is retrieved from the memory and proceeds to the output, \textbf{and}, within each rail, the control pulse experiences the photon addition operation\\ $\ket{\frac{\alpha}{\sqrt{2}}}\rightarrow(\hat{c}^\dagger)^n\ket{\frac{\alpha}{\sqrt{2}}}$. Due to the polarisation state of the signal, this induces the required phase between the creation operators acting on the rails of the control pulses, resulting in the transformation of the retrieval control pulse:
\begin{equation}
    \ket{\frac{\alpha}{\sqrt{2}}}_{c_u}
\ket{\frac{\alpha}{\sqrt{2}}}_{c_l}\rightarrow\left[\frac{(\hat{c}_u^\dagger-\hat{c}_l^\dagger)}{\sqrt{2}}\right]^n \ket{\frac{\alpha}{\sqrt{2}}}_{c_u}
\ket{\frac{\alpha}{\sqrt{2}}}_{c_l},
\end{equation}
where $\hat{c}_{u,l}^\dagger$ are the creation operators associated with the control in each of the two arms of the MZI.  This mirrors Eq.~(\ref{eqn:mzi_state_n}), thus showing that when this state passes through the final 50:50 beam splitter (BS), the dark port of the MZI will be populated with a Fock state $\ket{n}$. 

We have therefore implemented a transformation taking an input signal photon state $\ket{n}_\psi$, and the vacuum port of a balanced MZI and producing a 2-mode number entangled state:
\begin{equation}
    \ket{0}_D\ket{n}_\psi \rightarrow \ket{n}_D\ket{n}_\psi.
\end{equation}

Given the linear, bosonic nature of both the memory interaction (Eq. \ref{eq:full_mapping}) and interferometer transformation, the transformation holds term-by-term in the Fock basis. It therefore extends coherently to arbitrary superpositions, allowing us to generalise the transformation to, in the limit of perfect memory efficiency,
\begin{equation}
\sum_n a_n \ket{n}_s\ket{0}_D\rightarrow \sum_n a_n\ket{n}_s\ket{n}_D.
\label{eqn:final}
\end{equation}

\section{Applications}

\subsection{Non-destructive photon detection}

The transformation in Eq.~(\ref{eqn:final}) establishes photon-number correlations between the signal and auxiliary modes. This mapping directly enables quantum non-demolition measurement of photon number. A (destructive) measurement performed on the auxiliary mode infers the photon number of the signal without altering its occupation, while preserving coherence across Fock components prior to projection.

In contrast to current photonic QND schemes, the mechanism proposed here arises from a coherent, excitation-preserving mapping mediated by a quantum memory. Each excitation in the signal mode is independently converted into a correlated excitation in the auxiliary mode, ensuring a one-to-one correspondence in photon number. The interferometric mode selection isolates these excitations into a well-defined optical mode, enabling detection.

Photon number resolving measurements on the auxiliary mode commute with the photon-number operator of the signal mode, such that the measurement back-action is restricted to projection onto a number subspace. Repeated measurements would yield consistent outcomes, and the signal mode remains available for subsequent operations.

For finite memory efficiency $\eta < 1$, the dark-port measurement is a non-demolition measurement of the post-memory signal state rather than the input state. For an $n$-photon input, the probability that all excitations survive storage and retrieval scales as $\eta^{n}$, so partial loss can produce a detection of $k < n$ photons. The QND interpretation is therefore cleanest for single-photon inputs ($n = 1$), where a dark-port click unambiguously heralds the survival and presence of the signal photon. For $n > 1$, a photon-number-resolving measurement on the dark port projects the signal onto $|k\rangle_s$ and faithfully reports the transmitted photon number, with full QND property recovered in the limit $\eta \to 1$.

\subsection{Poisson Twin-Number States}

For a coherent-state signal input $|\alpha_s\rangle$, the transformation of Eq.~(\ref{eqn:final}) and the output of the QMI produces the two-mode entangled state
\begin{equation}
    \ket{\mathrm{PTNS}}= e^{-|\alpha_s|^2/2}\sum_n^\infty \frac{\alpha_s^n}{\sqrt{n!}}\ket{n}_s\ket{n}_D,
\end{equation}
which we term a \textit{Poisson Twin-Number State} (PTNS). This is a Poisson-weighted superposition of twin-Fock states $|n\rangle_s|n\rangle_D$ across two distinct modes.

The PTNS exhibits several distinctive properties. It possesses perfect photon-number correlations while each mode individually displays a Poissonian photon-number distribution with mean $|\alpha_s|^2$. This combination distinguishes the PTNS from the two-mode squeezed vacuum state $|\mathrm{TMSV}\rangle = \sqrt{1-\lambda^2}\sum_n \lambda^n |n\rangle|n\rangle$, which also has perfect number correlations but whose marginal distributions are thermal rather than Poissonian. 

The PTNS also has a simple Schmidt decomposition, with Schmidt coefficients 
\begin{equation} \lambda_n = e^{-|\alpha_s|^2/2} \frac{|\alpha_s|^n}{\sqrt{n!}}, 
\end{equation} 
and corresponding probabilities 
\begin{equation} p_n = |\lambda_n|^2 = e^{-|\alpha_s|^2} \frac{|\alpha_s|^{2n}}{n!}. 
\end{equation} 
Each mode therefore has mean photon number \begin{equation} \langle \hat n_s \rangle = \langle \hat n_D \rangle = |\alpha_s|^2, 
\end{equation} 
while the photon-number difference is perfectly defined, 
\begin{equation} (\hat n_s-\hat n_D)\ket{\mathrm{PTNS}} = 0, \end{equation} 
implying \begin{equation} \mathrm{Var}(\hat n_s-\hat n_D)=0. \end{equation} 
The PTNS is therefore an eigenstate of the photon-number difference operator with perfect number correlation between the two modes. As a pure bipartite state, its entanglement is quantified by the von Neumann entropy of either reduced state, 
\begin{equation} S_{\mathrm{PTNS}} = -\sum_{n=0}^{\infty} p_n \log p_n, \label{eq:PTNS_entropy} \end{equation} 
which is equal to the Shannon entropy of the associated Poisson distribution. The entanglement therefore increases monotonically with the coherent-state amplitude $|\alpha_s|$ and can, in principle, become arbitrarily large.

The PTNS is a non-Gaussian state and the quantum memory interferometer presented here provides, to our knowledge, the first direct mechanism for generating such a state from a coherent-state input while preserving the Poissonian photon-number statistics of the original field.

Non-Gaussian entangled states are widely recognised as essential resources for continuous-variable quantum information processing \cite{menicucci2006universal,weedbrook2012gaussian}. The PTNS is a new addition to this class, and characterising its utility for tasks such as entanglement distillation, error correction, and quantum sensing is an open question for future investigation.

Production of the PTNS is robust to finite memory efficiency. For a coherent-state input, a memory
with end-to-end efficiency $\eta$ acts as a lossy channel that maps $\ket{\alpha_s} \to \ket{\sqrt{\eta}\,\alpha_s}$,
preserving the coherent-state character of the input. The QMI transformation with this attenuated state produces
\begin{equation}
    \ket{\mathrm{PTNS}(\eta)} = e^{\frac{-\eta|\alpha_s|^2}{2}}\sum_{n=0}^{\infty}
    \frac{(\sqrt{\eta}\,\alpha_s)^n}{\sqrt{n!}}\ket{n}_s\ket{n}_D,
\end{equation}
which is simply a PTNS with effective amplitude $\alpha_{\mathrm{eff}} = \sqrt{\eta}\,\alpha_s$. The
output state remains pure and retains all of the entanglement and correlation properties of the
ideal PTNS; finite efficiency reduces only the effective mean photon number
$\bar{n}_{\mathrm{eff}} = \eta|\alpha_s|^2$, not the structure of the state. 

Projective measurement of the auxiliary mode onto photon number $n$ conditionally prepares the signal mode in the Fock state $|n\rangle_s$, enabling heralded Fock-state generation from a coherent-state input without photon subtraction or conventional nonlinear processes. The conditional fidelity would be set by the performance of the photon-number-resolving detector.

\subsection{Heralded single photon frequency conversion}

The transformation described here also provides a mechanism for single-photon, heralded frequency conversion based on coherent mapping between distinct optical modes. Frequency conversion is a key capability for quantum networks, enabling interfacing between systems operating at incompatible wavelengths. Conventional approaches rely on nonlinear optical processes such as three- and four-wave mixing, which require strong pump fields and stringent phase-matching conditions, and are often limited by bandwidth and noise \cite{wang2023quantum,cohen2026compact}.

In the present scheme, the signal and auxiliary modes are naturally defined at different optical frequencies through the memory interaction in atomic levels.

This mapping, therefore, could enable heralded single-photon level frequency conversion through conditional operation. Detection of an excitation in one mode heralds the presence of a corresponding photon in the other frequency mode, while preserving the quantum state of the signal.

\section{Experimental Constraints}

Experimental performance is governed by the following constraints, whose quantitative performance budgets are detailed in the Appendix.

\noindent \textbf{Signal preparation fidelity:} The input signal must be prepared in the $|{-}\rangle = (|H\rangle - |V\rangle)/\sqrt{2}$ polarisation state so that the PBD maps it onto the antisymmetric spatial mode $(u - l)/\sqrt{2}$ of the interferometer. Imperfect preparation (a residual $|{+}\rangle$ component) reduces the fraction of the photon addition that appears in the antisymmetric mode and therefore reduces the probability of the excitation reaching the dark port. This degrades the scheme's heralding efficiency but does not introduce false signals.

\noindent\textbf{Dark port noise:} Any imbalance between the two arms of the control-beam MZI, arising from amplitude mismatch, differential loss, phase drift, or spatial/temporal mode mismatch, allows leakage of the strong coherent field into the nominally dark port. These requirements place stringent but well-understood demands on interferometric stability and mode matching. For experimental operation, we propose using the control as the upper ``rung" in the ladder quantum memory (for example in $^{87}$Rb with the $5\mathrm{S}_{1/2} \to 5\mathrm{P}_{3/2}$ signal at 780nm and $5\mathrm{P}_{3/2}\to 4\mathrm{D}_{5/2}$ at 1529nm). The atomic vapour is therefore transparent to the control field, ensuring that the control-beam MZI experiences no differential dispersion or absorption from the atoms. This eliminates a potential source of interferometric imbalance that would arise if both fields were near-resonant with the vapour. Similarly, dark counts on the single-photon detector at the dark port constitute an irreducible noise floor.

\noindent\textbf{Memory performance:} The impact of memory efficiency is described in detail in the appendix, but is shown to predominantly act as a scaling factor to heralding efficiency and amplitude of the PTNS. The lifetime of the ladder memory is not a limiting factor for this scheme, as the signal needs only be stored for the duration of a single control-pulse cycle.

Importantly, many of these constraints (memory efficiency, spin-wave decoherence, and imperfect signal preparation) affect only the success probability of the protocol rather than the fidelity of the conditional output state.

\section{Conclusion}

We have shown that embedding a ladder-type quantum memory within a Mach--Zehnder
interferometer produces a coherent transformation that maps a single-mode photonic states
into number-correlated two-mode entangled states:
$\sum_n a_n \ket{n} \to \sum_n a_n \ket{n}_s\ket{n}_D$. This transformation arises from the combination of excitation-preserving photon addition by a quantum memory and interferometric mode selection, and we have designed an example optical configuration to achieve this in a Quantum Memory Interferometer (QMI).

The underlying mechanism supports several applications within a single architecture. For single-photon inputs, the QMI enables non-demolition detection: a click on the dark port unambiguously heralds the presence and survival of the signal photon. For coherent-state inputs, the QMI deterministically produces a Poisson Twin-Number State (PTNS). 

The results establish the QMI as a new platform for measurement-induced nonlinear photonics, in which effective photon--photon interactions are mediated by a light-matter interface rather than by optical nonlinearity. This opens a route to generating, characterising, and exploiting entangled photonic states, whilst noting that an experimental realisation of this device will require stringent interferometric stability and noise characterisation.

\section*{Acknowledgments}
I.F. acknowledges support from the EPSRC Hub in Quantum Computing and Simulation (EP/T001062/1). L.E.-P. and M.K. acknowledge support from EP/W524396/1.

\section*{Author Contributions} 

T.M.H. conceived the quantum memory interferometer and identified its applications to non-destructive photon detection, state preparation, and frequency conversion. S.B. contributed to early refinement of the idea. I.F. and V.B. assisted in the theoretical treatment and verified the analytical calculations presented in this work. L.E.-P. verified the approximations used, and M.K. considered the experimental implementation. T.M.H. supervised the project. All authors provided scientific input, critical discussion, and feedback on the theoretical development and manuscript. 
\bibliographystyle{apsrev4-2}
\bibliography{apssamp}

\clearpage \onecolumngrid \setcounter{section}{0} \setcounter{equation}{0} \setcounter{figure}{0} \setcounter{table}{0} \renewcommand{\thesection}{A} \renewcommand{\theequation}{A\arabic{equation}} \renewcommand{\thefigure}{A\arabic{figure}} \renewcommand{\thetable}{A\arabic{table}}

\begin{center} 
{\large \textbf{Supplementary Information: Number-correlated photon entanglement with a quantum memory interferometer}} 
\end{center} 
\section{Finite Efficiency}

We model finite memory efficiency $\eta$ as a beam-splitter interaction coupling the signal mode to an environmental loss mode. For an input Fock state $\ket{n}_s$, the lossy memory maps
\begin{equation}
    \ket{n}_s\ket{0}_E \;\longrightarrow\; \sum_{k=0}^{n}
    \sqrt{\binom{n}{k}}\,\eta^{k/2}(1-\eta)^{(n-k)/2}\,
    \ket{k}_s\ket{n-k}_E,
    \label{eq:loss-fock}
\end{equation}
where $k$ excitations survive the storage--retrieval cycle and $n - k$ are lost to the environment $E$. The QMI transformation acts on the surviving excitations, producing $\ket{k}_s\ket{k}_D$ in the signal and dark-port modes respectively. The full output state for a $n$ Fock-state input is therefore
\begin{equation}
    \ket{n}_s\ket{0}_D\ket{0}_E \;\longrightarrow\;
    \sum_{k=0}^{n} \sqrt{\binom{n}{k}}\,\eta^{k/2}(1-\eta)^{(n-k)/2}\,
    \ket{k}_s\ket{k}_D\ket{n-k}_E.
    \label{eq:qmi-loss-fock}
\end{equation}
For a general input state $\ket{\psi} = \sum_n a_n\ket{n}_s$, linearity gives
\begin{equation}
    \sum_n a_n\ket{n}_s\ket{0}_D\ket{0}_E \;\longrightarrow\;
    \sum_n a_n \sum_{k=0}^{n} \sqrt{\binom{n}{k}}\,
    \eta^{k/2}(1-\eta)^{(n-k)/2}\,
    \ket{k}_s\ket{k}_D\ket{n-k}_E.
    \label{eq:qmi-loss-general}
\end{equation}

Tracing over the environment yields the reduced density matrix for the signal and dark-port modes. Grouping terms by the number of lost excitations $\ell = n - k$, we obtain the Kraus decomposition \begin{equation}
    \hat{\rho}_{sD} = \sum_{\ell=0}^{\infty}
    \ket{\phi_\ell}\!\bra{\phi_\ell},
    \label{eq:kraus}
\end{equation}
where
\begin{equation}
    \ket{\phi_\ell} = (1-\eta)^{\ell/2}\sum_{k=0}^{\infty}
    a_{k+\ell}\,\sqrt{\binom{k+\ell}{k}}\;\eta^{k/2}\,
    \ket{k}_s\ket{k}_D.
    \label{eq:kraus-ops}
\end{equation}
Each Kraus operator $\ket{\phi_\ell}$ corresponds to exactly $\ell$ excitations lost to the environment. For a definite photon-number input $\ket{n}$, only a single value of $k$ contributes for each $\ell$, and the reduced state
\begin{equation}
    \hat{\rho}_{sD}^{(n)} = \sum_{k=0}^{n}\binom{n}{k}\,
    \eta^{k}(1-\eta)^{n-k}\,
    \ket{k}\!\bra{k}_s\otimes\ket{k}\!\bra{k}_D
    \label{eq:rho-fock}
\end{equation}
is a mixed state, a binomial distribution over twin-Fock components, with the outcome $\ket{k}_s\ket{k}_D$ occurring with probability $P(k)=\binom{n}{k}\eta^k(1-\eta)^{n-k}$.

\subsection{Coherent-state input.}
For a coherent-state input $\ket{\alpha}$, with $a_n = \alpha^n\frac{e^{-\frac{|\alpha|^2}{2}}}{\sqrt{n!}}$, the Kraus operators of Eq.~\eqref{eq:kraus-ops} become
\begin{equation}
    \ket{\phi_\ell} = e^{-\frac{|\alpha|^2}{2}}\,
    \frac{\bigl[\alpha\sqrt{1-\eta}\,\bigr]^\ell}{\sqrt{\ell!}}
    \;\sum_{k=0}^{\infty}
    \frac{\bigl[\alpha\sqrt{\eta}\,\bigr]^k}{\sqrt{k!}}\,
    \ket{k}_s\ket{k}_D.
    \label{eq:kraus-coherent}
\end{equation}
Every $\ket{\phi_\ell}$ is proportional to the same state vector in the signal--dark-port Hilbert space; the Kraus operators differ only by the scalar prefactor $[\alpha\sqrt{1-\eta}]^\ell/\sqrt{\ell!}$. It therefore follows that the reduced density matrix is pure:
\begin{equation}
    \hat{\rho}_{sD} = \ket{\mathrm{PTNS}(\sqrt{\eta}\,\alpha)}
    \!\bra{\mathrm{PTNS}(\sqrt{\eta}\,\alpha)},
    \label{eq:ptns-eta}
\end{equation}
where
\begin{equation}
    \ket{\mathrm{PTNS}(\sqrt{\eta}\,\alpha)} =
    e^{-\eta|\alpha|^2/2}\sum_{n=0}^{\infty}
    \frac{(\sqrt{\eta}\,\alpha)^n}{\sqrt{n!}}\,\ket{n}_s\ket{n}_D.
    \label{eq:ptns-eta-explicit}
\end{equation}
This is exactly the same transformation that yields a pure coherent state after a beam-splitter interaction, and hence the output of the QMI is simply a PTNS with effective amplitude $\alpha_{\mathrm{eff}} = \sqrt{\eta}\,\alpha$. The output state remains pure and retains all of the entanglement and correlation properties of the ideal PTNS. Finite efficiency reduces only the effective mean photon number $\bar{n}_{\mathrm{eff}} = \eta|\alpha|^2$, not the structure or purity of the state.

\subsection{Imperfect signal preparation}
\label{app:preparation}

We parametrise imperfect signal preparation by writing the polarisation state as
\begin{equation}
    \ket{\psi_{\mathrm{sig}}} = \cos(\theta)\,\ket{-} + \sin(\theta)\,\ket{+},
    \label{eq:imperfect-prep}
\end{equation}
where $(\theta) = 0$ corresponds to ideal preparation in the $\ket{-}$ state. The polarisation beam displacer maps $\ket{-}$ and $\ket{+}$ onto the antisymmetric and symmetric spatial modes of the interferometer, which after the final beamsplitter correspond to the dark-port and bright-port modes respectively.

An $n$-photon Fock state in this polarisation decomposes as
\begin{equation}
    \ket{n}_\psi = \sum_{k=0}^{n} \sqrt{\binom{n}{k}}\,
    \cos^k(\theta)\,\sin^{n-k}(\theta)\;\ket{k}_D\,\ket{n-k}_C,
    \label{eq:prep-decomp}
\end{equation}
where $\ket{k}_D$ denotes $k$ photons in the antisymmetric (dark-port) mode and $\ket{n-k}_C$ denotes $n-k$ photons in the symmetric (bright-port) mode.

After the QMI transformation, each signal excitation in the antisymmetric mode produces a corresponding control excitation at the dark port, while excitations in the symmetric mode are directed to the bright port and absorbed into the strong coherent field. Since the memory interaction preserves the photon number in each branch independently (implementing a number-shift rather than the bosonic creation operator [11]) the normalisation of each term is unity, and no dependence on the control-field amplitude $|\alpha|^2$ enters. The probability of detecting $k$ photons at the dark port, for an $n$-photon input with preparation error $(\theta)$, is
\begin{equation}
    P(k \,|\, n, (\theta)) = \binom{n}{k}\cos^{2k}(\theta)\,\sin^{2(n-k)}(\theta).
    \label{eq:prep-binom}
\end{equation}

Including finite memory efficiency $\eta$, each of the $k$ antisymmetric-mode excitations independently survives the storage--retrieval cycle with probability $\eta$. The compound distribution over preparation error and loss simplifies to a single binomial,
\begin{equation}
    P(j \,|\, n, (\theta), \eta) = \binom{n}{j}
    \bigl(\eta\cos^2(\theta)\bigr)^{\!j}\,
    \bigl(1-\eta\cos^2(\theta)\bigr)^{n-j},
    \label{eq:prep-eff}
\end{equation}
where $j$ is the number of detected dark-port photons. Preparation error and memory loss therefore enter on an equal footing, combining into a single effective parameter
\begin{equation}
    \eta_{\mathrm{eff}} = \eta\cos^2(\theta).
    \label{eq:eta-eff}
\end{equation}
Hence, the impact of imperfect signal preparation on successful heralding probability of a single input photon is given by $P(1)=\eta\cos^2(\theta)$. This result is independent of the control-field amplitude $|\alpha|^2$. Imperfect preparation reduces the heralding rate but does not couple to the intensity of the coherent field driving the interferometer. Excitations in the symmetric mode are directed entirely to the bright port and do not contribute noise at the dark port.

\section{Interferometric noise budget}
\label{app:noise}

The fidelity of the QMI transformation is ultimately limited by leakage of the
strong coherent control field into the dark output port. In this appendix
we quantify each noise contribution and derive the interferometric stability
requirements for single-photon--level operation.

We take as representative parameters a control pulse containing
$|\alpha|^2 = 10^{9}$ photons with a temporal pulse width
$T_p \approx 500~\mathrm{ps}$ (spectral bandwidth $\sim$2~GHz). These
values are consistent with the pulse energies used in current ORCA
implementations \cite{kaczmarek2018high,finkelstein2018fast}. We require the total spurious dark-port occupation to
satisfy $N_D^{(\mathrm{tot})} \ll 10^{-3}$ photons per pulse, ensuring that
false counts remain well below the single-photon herald signal.

\subsection{Phase noise}

Let $\delta\phi$ denote small fluctuations of the relative phase between the two
arms of the interferometer. To first order in $\delta\phi$, the dark-port field
amplitude and photon number become
\begin{equation}
    E_D^{(\phi)} \simeq \frac{\alpha\,\delta\phi}{2}\,,
    \qquad
    N_D^{(\phi)} = \frac{|\alpha|^2}{4}\,\delta\phi^2.
\end{equation}
Requiring $N_D^{(\phi)} < 10^{-3}$ photons per pulse gives
\begin{equation}
    \delta\phi < 2 \times 10^{-6}~\mathrm{rad}.
    \label{eq:phase-req}
\end{equation}
Environmental phase fluctuations typically occur on timescales
$\gtrsim 10~\mathrm{ns}$. Since $T_p = 500~\mathrm{ps}$ is much shorter than
these fluctuation times, the interferometer phase is effectively \emph{static}
over each pulse. Phase noise therefore manifests as slow, shot-to-shot drift
rather than intra-pulse dephasing, and can be suppressed by active stabilisation
operating in the kHz--MHz regime. Operating the control field at a telecom wavelength far-detuned from atomic
resonance renders the vapour transparent to the control beam, eliminating
atom-induced differential phase shifts between the interferometer arms.

\subsection{Amplitude imbalance}

Let the field amplitudes in the two arms be $\alpha(1+\epsilon_a)$ and $\alpha$,
where $\epsilon_a \ll 1$ parameterises a fractional imbalance arising from
differential loss or beamsplitter imperfection. The resulting dark-port leakage is
\begin{equation}
    E_D^{(\mathrm{amp})} = \frac{\alpha\,\epsilon_a}{2}\,,
    \qquad
    N_D^{(\mathrm{amp})}
    = \frac{|\alpha|^2}{4}\,\epsilon_a^2.
\end{equation}
Demanding $N_D^{(\mathrm{amp})} < 10^{-3}$ yields
\begin{equation}
    \epsilon_a < 2 \times 10^{-6}.
    \label{eq:amp-req}
\end{equation}

\subsection{Spatial and temporal mode mismatch}

Imperfect spatial or temporal overlap between the two arms is parameterised by
the mode overlap $1 - \mu$, where $\mu \ll 1$ quantifies the fractional component
of one arm that is orthogonal to the other. The orthogonal component does not
interfere destructively and leaks directly into the dark port:
\begin{equation}
    E_D^{(\mathrm{mode})} \simeq \alpha\,\mu\,,
    \qquad
    N_D^{(\mathrm{mode})} = |\alpha|^2\,\mu^2.
\end{equation}
Note that mode mismatch enters without the factor of $1/4$ present in phase and
amplitude noise, making it the most sensitive parameter. To maintain
$N_D^{(\mathrm{mode})} < 10^{-3}$,
\begin{equation}
    \mu < 10^{-6}.
    \label{eq:mode-req}
\end{equation}
For Gaussian pulses, a timing offset $\Delta t$ between the two arms produces
$\mu \approx (\Delta t / T_p)^2$ for small $\Delta t$. Maintaining
$\mu < 10^{-6}$ therefore requires $\Delta t < 10^{-3}\,T_p \approx 0.5~\mathrm{ps}$,
which is readily achievable with standard optical path-length matching.

\subsection{Detector dark counts}

Let $p_{\mathrm{dc}}$ denote the per-pulse dark-count probability of the
single-photon detector monitoring the dark port. To ensure that detector noise
remains below the herald signal, we require
\begin{equation}
    p_{\mathrm{dc}} \ll 10^{-3}.
\end{equation}
Superconducting nanowire single-photon detectors (SNSPDs) gated to a
100--500~ps window achieve $p_{\mathrm{dc}} \sim 10^{-7}$ per gate, satisfying
this requirement by four orders of magnitude.

\subsection{Total dark-port noise}

Summing all contributions, the total spurious dark-port photon number per pulse
is
\begin{equation}
    N_D^{(\mathrm{tot})}
    =  N_D^{(\phi)}
    + N_D^{(\mathrm{amp})}
    + N_D^{(\mathrm{mode})}
    + p_{\mathrm{dc}}
    \ll 10^{-3}.
    \label{eq:total-noise}
\end{equation}
The hierarchy of contributions is summarised in Table~\ref{tab:noise}.

\begin{table}[h]
\centering
\caption{Dark-port noise budget for $|\alpha|^2 = 10^{9}$ and
$T_p = 500~\mathrm{ps}$, assuming each imperfection is held at its
threshold value.}
\label{tab:noise}
\begin{tabular}{lcc}
\hline\hline
Source & Requirement & $N_D$ at threshold \\
\hline
Phase noise         & $\delta\phi < 2\times 10^{-6}$~rad       & $10^{-3}$ \\
Amplitude imbalance & $\epsilon_a < 2\times 10^{-6}$           & $10^{-3}$ \\
Mode mismatch       & $\mu < 10^{-6}$                          & $10^{-3}$ \\
Detector dark counts& $p_{\mathrm{dc}} < 10^{-3}$             & $\sim 10^{-7}$ \\
\hline
\textbf{Total}      &                                          & $\sim 10^{-3}$ \\
\hline\hline
\end{tabular}
\end{table}

The dominant noise mechanisms are phase drift, amplitude imbalance, and mode
mismatch, each requiring control at the part-per-million level. Shot-noise
leakage and detector dark counts are negligible by comparison. Crucially, the
symmetry of the balanced MZI ensures that all classical noise from the strong
coherent field occupies the symmetric spatial mode and is directed to the bright
port, while the herald signal from the quantum memory populates the antisymmetric
mode exclusively. Dark-port contamination arises only from deviations from
perfect balance.

We note that the noise budget is independent of the quantum memory interaction:
the ORCA ladder-type scheme has been shown to introduce no noise photons into the signal, so the interferometric imperfections described here constitute a complete noise model for the scheme.

\section{Shifted Coherent State Approximation}

The addition of $n$ photons to a coherent state is defined by
\begin{equation}
(\hat{c}^{\dagger})^n\lvert\alpha\rangle
=e^{-{\lvert \alpha \rvert ^2}/{2}}
\sum_{N=0}^{\infty}
\frac{\alpha^N}{\sqrt{N!}}(\hat{c}^{\dagger})^n\lvert N\rangle.
\label{eq:photon_added_coherent_state}
\end{equation}
To expand this expression, we consider $n$ applications of the creation operator on a Fock state $\lvert N \rangle$
\begin{equation}
(\hat{c}^{\dagger})^n\lvert N\rangle = \sqrt{\frac{(N+n)!}{N!}}\lvert N+n\rangle,
\end{equation}
which, when substituted into Eq.~(\ref{eq:photon_added_coherent_state}), gives 
\begin{equation}
(\hat{c}^{\dagger})^n\lvert\alpha\rangle
=
e^{-{\lvert \alpha \rvert ^2}/{2}}
\sum_{N=0}^{\infty}
\frac{\alpha^N}{\sqrt{N!}}\sqrt{\frac{(N+n)!}{N!}}\lvert N+n\rangle.
\end{equation}
The normalisation factor $A_n$ such that $\langle N \rvert c^n A_n^*A_n(c^{\dagger})^n\lvert N\rangle = 1$ is given by
\begin{equation}
A_n = \left[ e^{-\lvert \alpha \rvert ^2}
\sum_{N=0}^{\infty}
\frac{\lvert \alpha \rvert ^{2N}}{{N!}}{\frac{(N+n)!}{N!}} \right] ^{-\frac{1}{2}} 
= \frac{1}{\sqrt{{\langle g(N,n)^2 \rangle}}}, 
\end{equation}
where in the final step we have defined $g(N,n)=\sqrt{{(N+n)!}/{N!}}$, and recognise that the normalisation factor is the expected value of $g(N)$ given that photon number $N$ follows a Poisson distribution. Therefore, the normalised $n$-photon-added coherent state can be written as
\begin{equation}
A_n(\hat{c}^{\dagger})^n\lvert N\rangle = 
\frac{e^{-{\lvert \alpha \rvert ^2}/{2}}}{\sqrt{\langle g(N,n)^2 \rangle}} 
\sum_{N=0}^{\infty}
\frac{\alpha^N}{\sqrt{N!}}g(N,n)\lvert N+n\rangle.
\label{eqn:norm-added}
\end{equation}

In order to validate the approximation between Eq.~\ref{eqn:norm-added} and the result of the number shift operator on a coherent state $\hat S_n |\alpha\rangle$, we use state fidelity to define the degree of approximation. The fidelity $F(N,n)$ is expressed in terms of the expectation values of the function $g(N,n)$ as 
\begin{equation}
F(N,n) = \lvert \langle \alpha|\hat{S}_n A_n (c^\dagger)^n |\alpha \rangle \rvert^2 
= \left[ \frac{e^{-\lvert\alpha\rvert^2}}{\sqrt{\langle g(N,n)^2 \rangle}}
\sum_{N=0}^{\infty} \frac{\lvert\alpha\rvert^{2N}}{{N!}}g(N,n) \right]^2 
= \frac{\langle g(N,n) \rangle ^2}{\langle g(N,n)^2 \rangle}.
\label{eq:F(N,n)}
\end{equation}
To obtain an analytical expression for the fidelity, the expectation values are evaluated using Poisson moments, necessitating Taylor series expansion and approximations to simplify expectation values that do not have a well-defined moment, such as $\langle g(N,n) \rangle$. We note that we expect $n$ to be $\mathcal{O}(1)$ and $\lvert\alpha\rvert^2$ to be $\mathcal{O}(10^9)$ in practical implementations, and therefore we make approximations to first order in $\lvert\alpha\rvert^{-2}$. Additionally, as $\lvert\alpha\rvert^2$ becomes sufficiently large, the Poisson distribution of the photon number $N$ will become increasingly concentrated around the mean, and therefore $N \approx \lvert\alpha\rvert^2$. As a result, we make approximations to first order in $N^{-1}$ where necessary. In the following, we explicitly derive an analytical expression for $\langle g(N,n) \rangle$, and then state the result for $\langle g(N,n)^2 \rangle$ given that it follows from a similar methodology. 

We begin by rewriting the factorial expression 
\begin{equation}
g(N,n) = \sqrt{\frac{(N+n)!}{N!}} = N^{\frac{n}{2}}\prod_{k=1}^n (1+\frac{k}{N})^{\frac{1}{2}}.
\end{equation}
We use a logarithm to transform the product, and substitute the Taylor expansion of $\ln(1+x)$, giving
\begin{equation}
\ln(g(N,n)) = \frac{n}{2}\ln(N) + \frac{1}{2}\sum_{k=1}^n\ln(1+\frac{k}{N}) = \frac{n}{2}\ln(N) + \frac{1}{2}\sum_{k=1}^n (\frac{k}{N} + \mathcal{O}(N^{-2}))=\frac{n}{2}\ln(N)+\frac{n(n+1)}{4N},
\end{equation}
where in the final step, the Taylor expansion is truncated at $\mathcal{O}(N^{-1})$. We then resolve the logarithm, giving the final expression for $g(N,n)$ as
\begin{equation}
g(N,n) = N^{\frac{n}{2}} \exp(\frac{n(n+1)}{4N}) = N^{\frac{n}{2}} (1+\frac{n(n+1)}{4N}+\mathcal{O}(N^{-2})),
\label{eq:g(N,n)}
\end{equation}
where the Taylor expansion of the exponential has been truncated at $\mathcal{O}(N^{-1})$. This form of the function $g(N,n)$ enables an analytical expression for the expectation value through Poisson moments. To do this, we rewrite the photon number $N$ in terms of a deviation from the mean,
\begin{equation}
N = \lvert\alpha\rvert^2 + \delta N
\label{eq:mean_deviation}
\end{equation}
where $\lvert\alpha\rvert^2$ is the mean photon number for the Poisson distribution of a coherent state and $\delta N =(N-\lvert\alpha\rvert^2)$. The Poisson central moments are identified as $\langle \delta N \rangle=0$ and $\langle (\delta N)^2 \rangle=\lvert\alpha\rvert^2$. Substituting Eq.~(\ref{eq:mean_deviation}) into Eq.~(\ref{eq:g(N,n)}) gives
\begin{equation}
g(\lvert\alpha\rvert ^2, \delta N, n)
=(\lvert\alpha\rvert^2 + \delta N)^{\frac{n}{2}} + \frac{n(n+1)}{4}(\lvert\alpha\rvert^2 + \delta N)^{\frac{n}{2}-1} 
= \lvert\alpha\rvert^n(1+\frac{\delta N}{\lvert\alpha\rvert^2})^{\frac{n}{2}}+\lvert\alpha\rvert^{(n-2)}\frac{n(n+1)}{4}(1+\frac{\delta N}{\lvert\alpha\rvert^2})^{\frac{n}{2}-1}.
\end{equation}
As ${\delta N}/{\lvert\alpha\rvert^2} <<1$, we use a second order binomial approximation to expand the exponents giving
\begin{equation}
g(\lvert\alpha\rvert^2, \delta N, n)
=\lvert\alpha\rvert^n(1+\frac{\delta N}{\lvert\alpha\rvert^2}\frac{n}{2} + \frac{(\delta N)^2}{(\lvert\alpha\rvert^2)^2}\frac{n(n-2)}{8}) 
+ \lvert\alpha\rvert^{n-2}\frac{n(n+1)}{4}(1+\frac{\delta N}{\lvert\alpha\rvert^2}(\frac{n}{2}-1)
+\frac{(\delta N)^2}{(\lvert\alpha\rvert^2)^2}\frac{(n-2)(n-4)}{8}).
\end{equation}
We then evaluate the expectation value up to $\mathcal{O}(\lvert \alpha\rvert^{-2})$, resulting in the final expression
\begin{equation}
\langle g(\lvert\alpha\rvert^2, n) \rangle 
= \lvert\alpha\rvert^n (1 + \frac{n(n-2)}{8\lvert\alpha\rvert^2} + \frac{n(n+1)}{4\lvert\alpha\rvert^2}) 
= \lvert\alpha\rvert^n(1+\frac{3n^2}{8\lvert\alpha\rvert^2}).
\end{equation}

A similar argument is followed to evaluate the expectation value $\langle g(N,n)^2 \rangle$ under consistent approximations. The resulting expression is given by
\begin{equation}
\langle g(N,n)^2 \rangle \rightarrow \langle g(\lvert\alpha\rvert^2,n)^2 \rangle 
= \lvert\alpha\rvert^{2n}(1+\frac{n^2}{\lvert\alpha\rvert^2}).
\end{equation}
This enables a final expression for fidelity as a function of added photon number and mean coherent state amplitude. Returning to the fidelity equation given in Eq.~(\ref{eq:F(N,n)}), we substitute the resulting expressions for $\langle g(\lvert\alpha\rvert^2,n)^2 \rangle$ and $\langle g(\lvert\alpha\rvert^2, n) \rangle$, giving 
\begin{equation}
F(\lvert\alpha\rvert^2, n) 
= (1+\frac{3n^2}{4\lvert\alpha\rvert^2})^2(1+\frac{n^2}{\lvert\alpha\rvert^2})^{-1} 
= 1-\frac{n^2}{4\lvert\alpha\rvert^2},
\end{equation}
where in the final step, a binomial expansion has been used to $\mathcal{O}(\lvert \alpha\rvert^{-2})$ to simplify the exponents. We note that deviation from unit fidelity is minimised in the limit that $\lvert\alpha\rvert^2>>n^2$. In particular, fidelity values greater than $99.99 \%$ are achieved for $\lvert\alpha\rvert^2>10^5$ when the number of added photons is $\mathcal{O}(1)$. 

\begin{figure*}[ht]
    \centering
    \includegraphics[width=0.6\linewidth]{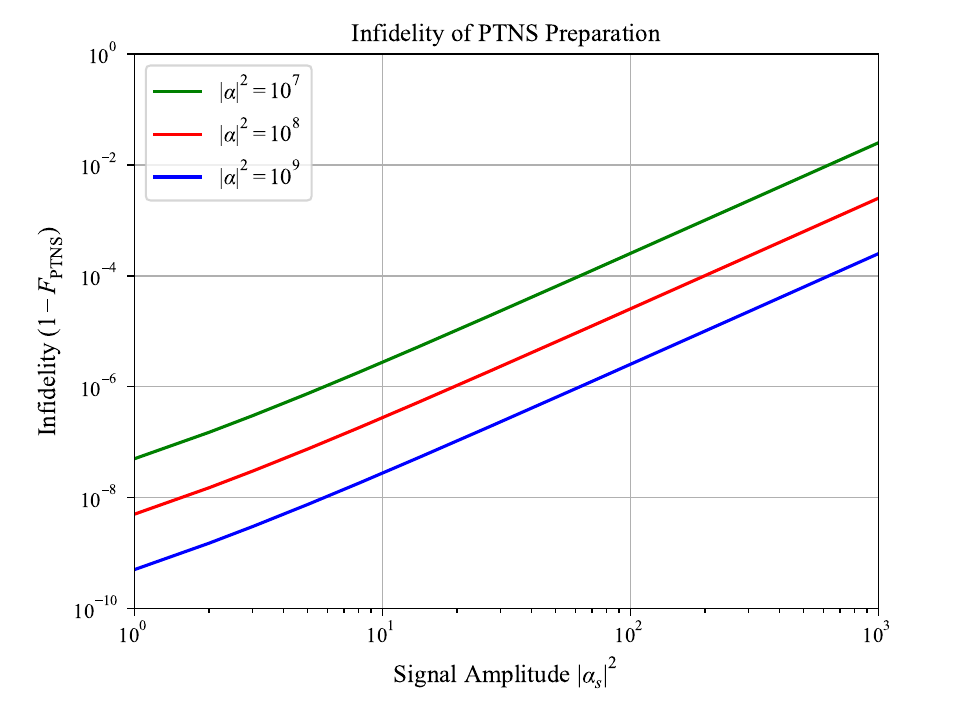}
    \caption{The infidelity ($1-F$) of the PTNS preparation. In reference to Eq. (\ref{eq:ptns_infidelity}), the infidelity depends upon the ratio of the signal and control field amplitudes. Experimentally, control field amplitudes $\lvert\alpha\rvert^2$ of $\mathcal{O}(10^9)$ are expected, for which the fidelity of the PTNS preparation sits above a $99.99 \%$ threshold for signal amplitudes $\lvert\alpha_s\rvert^2 \approx \mathcal{O}(10^2)$.}
    \label{fig:ptns_infidelity}
\end{figure*}

For instances where the input signal to the device is a superposition of number states, the fidelity of the approximation will be averaged according to the probability amplitudes within the superposition. This is particularly relevant for the preparation of the Poisson Twin Number State (PTNS), where the input signal is a coherent state with Poisson-distributed probability amplitudes. The fidelity of PTNS preparation $F_{\text{PTNS}}(\lvert \alpha \rvert ^2, \lvert \alpha_s \rvert ^2, n)$ is therefore given by the expectation value of $F(\lvert \alpha \rvert ^2, n)$ where $n$ follows a Poisson distribution with average photon number $\lvert \alpha_s\rvert^2$. The resulting fidelity expression is
\begin{equation}
F_{\text{PTNS}}(\lvert \alpha \rvert ^2, \lvert \alpha_s \rvert ^2, n)= \langle F(\lvert \alpha \rvert ^2, n) \rangle = e^{-\lvert \alpha_s \rvert ^2} 
\sum_{n=0}^{\infty}
\frac{\lvert \alpha_s \rvert ^{2n}}{n!}(1-\frac{n^2}{4\lvert \alpha \rvert ^2}) =
1-\frac{\lvert \alpha_s \rvert ^4+\lvert \alpha_s \rvert ^2}{4\lvert \alpha \rvert ^2},
\label{eq:ptns_infidelity}
\end{equation}
where the expectation value has been evaluated using the second raw moment of the Poisson distribution $\langle n^2\rangle = \lvert \alpha_s \rvert ^4+\lvert \alpha_s \rvert ^2$. Therefore, the deviation from perfect fidelity will depend upon the ratio of the mean photon numbers of the signal and control pulse. This infidelity ($1-F_{\text{PTNS}}$) is plotted in Fig. (\ref{fig:ptns_infidelity}) for varying control pulse amplitudes. We highlight that if $\lvert \alpha \rvert^2$ is  $\mathcal{O}(10^9)$, as is expected for practical implementation, then the fidelity remains above a $99.99 \%$ threshold for signal amplitudes of $\mathcal{O}(10^2)$ photons.

\end{document}